\documentclass[conference]{IEEEtran}
\IEEEoverridecommandlockouts
\usepackage{cite}
\usepackage{amsmath,amssymb,amsfonts}
\usepackage{algorithmic}
\usepackage{graphicx}
\usepackage{textcomp}
\usepackage{xcolor}
\usepackage{subfigure}
\usepackage[a4paper, total={184mm,239mm}]{geometry}
\usepackage{url}
\usepackage{tikz}
\newcommand{\circlednum}[1]{%
  \tikz[baseline=(char.base)]{
    \node[shape=circle,fill=black,inner sep=1pt] (char) {\textcolor{white}{#1}};
  }%
}
\usepackage{xcolor}
\newcommand{\code}[1]{\texttt{\textcolor{black!80!black}{#1}}}
\def\BibTeX{{\rm B\kern-.05em{\sc i\kern-.025em b}\kern-.08em
    T\kern-.1667em\lower.7ex\hbox{E}\kern-.125emX}}
\begin{document}

\newcommand{\KI}[1]{\textcolor{green}{#1}}

\newcommand{\SX}[1]{\textcolor{blue}{#1}}

\title{Multi-Partner Project: Multi-GPU Performance Portability Analysis for CFD Simulations at Scale\\
\thanks{This work has been partially funded by the EU Horizon 2022 program under grant
agreement No 101096698 REFMAP (website: https://www.refmap.eu).
We acknowledge the EuroHPC Joint Undertaking for awarding this project access to the EuroHPC supercomputer LUMI, hosted by CSC (Finland) and the LUMI consortium through a EuroHPC Regular Access call.}
}

\author{\IEEEauthorblockN{Panagiotis-Eleftherios
Eleftherakis\IEEEauthorrefmark{1},
George Anagnostopoulos\IEEEauthorrefmark{1},
Anastassis Kapetanakis\IEEEauthorrefmark{1},
Mohammad Umair\IEEEauthorrefmark{2}, \\
Jean-Yves Vet\IEEEauthorrefmark{3}, 
Konstantinos Iliakis\IEEEauthorrefmark{1},
Jonathan Vincent\IEEEauthorrefmark{2},
Jing Gong\IEEEauthorrefmark{2},
Akshay Patil\IEEEauthorrefmark{4}, 
Clara García-Sánchez\IEEEauthorrefmark{4}, \\ 
Gerardo Zampino\IEEEauthorrefmark{2},
Ricardo Vinuesa\IEEEauthorrefmark{5},
Sotirios Xydis\IEEEauthorrefmark{1}}
\and
\IEEEauthorblockA{\IEEEauthorrefmark{1}\textit{National Technical University of Athens, Greece}}
\and
\IEEEauthorblockA{\IEEEauthorrefmark{2}\textit{KTH Royal Institute of Technology, Sweden}}
\and
\IEEEauthorblockA{\IEEEauthorrefmark{3}\textit{Hewlett Packard Enterprise (HPE), France}}
\and
\IEEEauthorblockA{\IEEEauthorrefmark{4}\textit{Technical University of Delft, Netherlands,}}
\and
\IEEEauthorblockA{\IEEEauthorrefmark{5}\textit{University of Michigan, USA}}
}

\maketitle

\begin{abstract}
As heterogeneous supercomputing architectures leveraging GPUs become increasingly central to high-performance computing (HPC), it is crucial for computational fluid dynamics (CFD) simulations, a de-facto HPC workload, to efficiently utilize such hardware. One of the key challenges of HPC codes is performance portability, i.e. the ability to maintain near-optimal performance across different accelerators. In the context of the \textbf{REFMAP} project, which targets scalable, GPU-enabled multi-fidelity CFD for urban airflow prediction, this paper analyzes the performance portability of SOD2D, a state-of-the-art Spectral Elements simulation framework across AMD and NVIDIA GPU architectures.
We first discuss the physical and numerical models underlying SOD2D, highlighting its computational hotspots. Then, we examine its performance and scalability in a multi-level manner, i.e. defining and characterizing an extensive full-stack design space spanning across application, software and hardware infrastructure related parameters. Single-GPU performance characterization across server-grade NVIDIA and AMD GPU architectures and vendor-specific compiler stacks, show the potential as well as the diverse effect of memory access optimizations, i.e. 0.69$\times$ - 3.91$\times$ deviations in acceleration speedup. Performance variability of SOD2D at scale is further examined on the LUMI multi-GPU cluster, where profiling reveals similar throughput variations, highlighting the limits of performance projections and the need for multi-level, informed tuning.
\end{abstract}


\section{Introduction}
Computational Fluid Dynamics (CFD) is an essential tool for modern engineering, enabling detailed analysis of fluid behavior in aerospace, automotive, maritime, and environmental systems without costly physical prototypes. Recently, CFD-generated datasets have also become central to training physics-informed neural models, which embed physical laws into machine learning models for improved predictive accuracy~\cite{Vinuesa_2022, Eivazi_2022}.
At its core, CFD solves the Navier–Stokes equations, expressing conservation of mass, momentum, and energy in fluid flows. Analytical solutions are rare, so numerical methods such as the Finite Difference Method (FDM) and Finite Element Method (FEM) are employed. FDM is efficient for structured grids, while FEM better handles complex geometries at higher computational cost.

High-Performance Computing (HPC) has greatly advanced CFD, enabling high-fidelity simulations of complex phenomena. Heterogeneous architectures, particularly those leveraging Graphics Processing Units (GPUs), have driven progress toward exascale computing, supported by sophisticated solvers such as OpenFOAM, NekRS, and others~\cite{fischer2022nekrs, jasak2009openfoam}. 
However, optimizing CFD solvers for diverse hardware remains challenging due to differing GPU architectures, compiler ecosystems, and programming models. Solvers optimized for one platform often underperform on another, hindering performance portability—i.e., sustained efficiency across architectures.

Within this context, the EU funded \textbf{REFMAP} project (\textit{Reducing Environmental Footprint through transformative Multi-scale Aviation Planning}) targets low-latency, multi-fidelity urban-airflow prediction to support sustainable Innovative Air Mobility in urban environments. Its high-fidelity CFD component couples GPU-enabled simulations with machine-learning surrogates. High-resolution Direct Numerical Simulation (DNS) data train compact predictors deployable at the edge or in the cloud. Since such workloads are computationally intensive, REFMAP integrates GPU-accelerated CFD and autotuning to overcome data-generation bottlenecks. The Spectral Element solver \textbf{SOD2D}\cite{GASPARINO2024109067} is ported to GPUs using OpenACC \cite{openacc27} and MPI, enabling scalable, turbulence-resolving simulations across NVIDIA and AMD systems. Achieving performance portability across these heterogeneous HPC platforms is critical to REFMAP’s data-driven workflow, ensuring scalable, cost-effective surrogate training and deployment.

CFD solvers adopting high-order numerical schemes and multi-architecture support—such as URANOS~\cite{DEVANNA2024109285}, STREAmS~\cite{SATHYANARAYANA2025104993}, and PyFR~\cite{WITHERDEN20143028}—have achieved significant GPU speedups, though cross-vendor analysis remains limited. Reguly and Mudalige~\cite{REGULY2020104425} provide a broad overview of performance portability strategies, noting that few works (e.g., CloverLeaf~\cite{mcintosh2019performance}, TeaLeaf~\cite{tealeaf}, Nekbone~\cite{nekbone}) use OpenACC across multiple platforms, with limited direct cross-vendor comparison. More recent efforts ~\cite{DAI2024104831} similarly omit direct cross-vendor GPU evaluation. To our knowledge, this work is among the first to perform a detailed, full-stack cross-vendor study of GPU-accelerated CFD using OpenACC, spanning from solver configuration to hardware-level memory optimization.


\begin{figure*}[t]
  \centering
  \includegraphics[width=\textwidth]{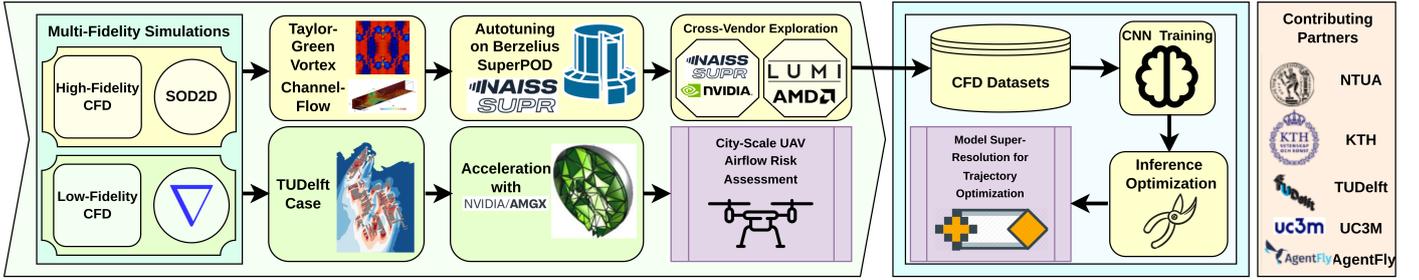}
  \caption{Overview of REFMAP's HPC to ML workflow, with emphasis on CFD acceleration.}
  \label{fig:refmap}
\end{figure*}

\section{The REFMAP Project Outline}
\label{sec:refmap-concept}

REFMAP targets low-latency, multi-fidelity urban-airflow prediction to support sustainable Innovative Air Mobility. As seen in Fig. \ref{fig:refmap}, the low-fidelity branch is driven by the OpenFOAM CFD simulator (and accelerated by the NVIDIA AmgX solver) executing a model that aims at evaluating the neighbourhood-scale wind-associated risk \cite{patil}. Regarding the high-fidelity framework, the SOD2D simulator \cite{GASPARINO2024109067} is employed (a Fortran-based SEM code with OpenACC offload and MPI partitioning, where key kernels —i.e. convection, diffusion— are parallelized across elements and nodes) enabling the coupling of GPU-enabled CFD with data-driven surrogates. High-resolution LES/DNS simulations generate turbulence-resolving data to train compact models \cite{guastoni_20} that undergo extensive optimizations such as compression and pruning in order to deploy at the edge or in the cloud, enabling UAV trajectory optimization and high-fidelity sensor placement, thus providing a robust substrate for efficient and safe UAV navigation. On CPU-only systems, such cases require weeks to compute, limiting the datasets for surrogate training. REFMAP overcomes this bottleneck by extensive autotuning and performance exploration of SOD2D. 
Initial studies have shown performance sensitivity when freely varying gang numbers and vector lengths using OpenTuner's AUC-Bandit search technique, obtaining 1.2--1.5\(\times\) speedup \cite{date24,samos_refmap}. Moving towards REFMAP's mature phase, multiple infrastructures were employed necessitating multi-level performance exploration \cite{cf_25} - i.e. spanning across hardware,software and application-specific parameters.
\textit{On the surrogate model side,} REFMAP employs fully convolutional networks (FCNs) that map wall-shear-stress and pressure inputs to velocity-fluctuation fields \(\hat{u}, \hat{v}, \hat{w}\) at selected wall-normal positions. Models are trained on DNS data (Re\(_\tau\)=180) with 6,000 snapshots per \(y^+=15, 30, 50, 100\), using MSE loss and normalized targets to balance component errors. While additional GAN and transformer models are explored, FCNs remain the main surrogate coupled with accelerated CFD for wall-based flow reconstruction. Leveraging structured pruning, quantization, and compiler-level optimizations, the final turbulence-prediction models achieve up to 50× compression with negligible loss in accuracy, significantly improving latency and throughput. These advances enable fast, reliable flow prediction and establish a foundation for integrating high-performance simulation and data-driven surrogates into real-time UAV navigation and urban risk assessment workflows. The rest of this paper will focus on the performance portability analysis that enables efficient generation of datasets for surrogate model training within the REFMAP workflow.

\section{SOD2D: Spectral FEM for compressible scale-resolving simulations}

\subsection{Implementation and governing equations}

The physical model considered is composed of the Navier–Stokes equations governing mass, momentum, and energy conservation in a compressible fluid \cite{GASPARINO2024109067}. In conservative form:

\begin{align}
&\textcolor[HTML]{006400}{\frac{\partial \rho}{\partial t}} 
\textcolor{orange}{+ \nabla \cdot (\rho \boldsymbol{u})} = 0 \label{eq:nv-mass} \\
&\textcolor[HTML]{006400}{\frac{\partial (\rho \boldsymbol{u})}{\partial t}} 
\textcolor{orange}{+ \nabla \cdot (\rho \boldsymbol{u} \otimes \boldsymbol{u})} 
\textcolor{blue}{- \nabla \boldsymbol{\tau}} = - \nabla p + \boldsymbol{f} \label{eq:nv-mom} \\
&\textcolor[HTML]{006400}{\frac{\partial E}{\partial t}} 
\textcolor{orange}{+ \nabla \cdot ((E+p)\boldsymbol{u})} 
\textcolor{blue}{- \nabla \cdot (\boldsymbol{\tau} \boldsymbol{u}) - \nabla (\kappa \nabla T)} = S \label{eq:nv-ener}
\end{align}

Here $\rho$, $\boldsymbol{u}$, $p$, $T$, and $E$ denote density, velocity, pressure, temperature, and total energy. The system is closed with the ideal gas law.  
The colorized terms highlight \textcolor[HTML]{006400}{\textbf{Temporal Evolution}}, \textcolor{orange}{\textbf{Convection}}, and \textcolor{blue}{\textbf{Diffusion}}.
\label{sec:numerical-model}
The equations are solved by separating temporal and spatial discretization. Spatial terms use the Finite Element Method (FEM) and temporal integration uses a fourth-order Runge–Kutta (RK4) scheme.

\label{sec:sod2d}
\begin{figure}
    \centering
    \includegraphics[width=1\linewidth]{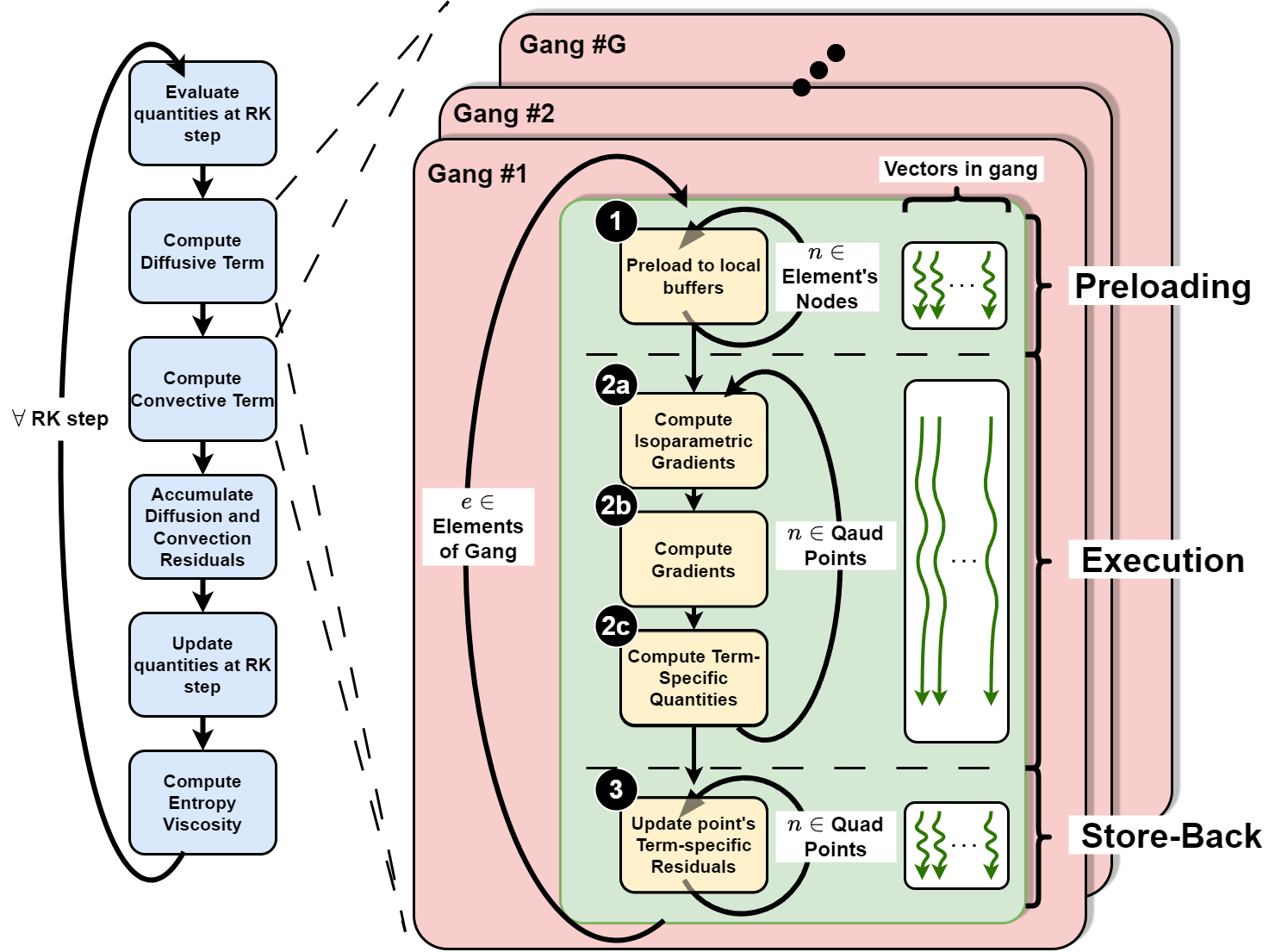}
    \caption{High-level algorithmic overview of SOD2D}
    \label{fig:SOD2D-flow}
    \vspace{-10px}
\end{figure}

SOD2D \cite{GASPARINO2024109067} is an OpenACC-based CFD software leveraging MPI process partitioning techniques,  thus targeting multiple platforms. It focuses on a variant of the Finite Element Method (FEM) known as the Spectral Element Method (SEM) and can handle unstructured meshes, thus modeling complex geometries. This method utilizes non-uniformly distributed nodes within each element while lowering the computational cost for high mesh resolutions. Additionally, the Gauss-Lobatto-Legendre quadrature rule is employed aligning the quadrature points with the element nodes in the discretized mesh.

The aforementioned design choices of SOD2D reflect its primary objective: to create a hardware-optimized CFD simulator capable of maximizing parallelism across multi-core CPUs as well as single- and multi-GPU architectures \cite{GASPARINO2024109067, date24}. The mathematical manipulation of the PDEs within the SEM framework results in a computationally efficient diagonal linear system. This structure eliminates the reliance on complex linear solvers, which are often a major bottleneck in traditional CFD applications~\cite{REGULY2020104425}. 


\subsection{SOD2D execution flow}

Fig.~\ref{fig:SOD2D-flow} illustrates the execution flow of SOD2D’s various computational kernels, represented by the blue rectangles. The algorithm is structured around a loop over four Runge-Kutta steps, as described in \cite{GASPARINO2024109067}. Within each step, the necessary values for evaluating the Runge-Kutta (RK) function $\boldsymbol{f}$ are computed. This is followed by the evaluation of the Diffusive and Convective term residuals at each point $n$ within every mesh element $e$. Subsequently, the residuals of these two terms are accumulated at each node and the quantities' values are updated at the current RK step. Finally, entropy viscosity is computed, a numerical technique used to stabilize fluid flow simulations, particularly in shock-resolving scenarios
 
The focus lies on the computation of the Diffusive and Convective terms, a process divided into three subparts: Preloading, Execution and Store-Back. These subparts are executed independently, with synchronization barriers indicated by dashed lines. Each term computation iterates over all elements in the mesh. This element loop is distributed across fully parallel gangs, and vectors within gangs, as defined in \cite{openacc27}, each with sizes defined by the kernel’s execution parameters.
In step \circlednum{1}, global quantities corresponding to the nodes of the current element are fetched into local arrays, thus the main execution part begins. The first operation involves calculating the isoparametric gradients \circlednum{2a}, which are gradients in the natural coordinate system. Those are then used to compute the gradients in the global coordinate system \circlednum{2b} by multiplying them with the inverse \textit{Jacobian}. In the next step \circlednum{2c}, additional term-specific quantities are computed. This process is repeated for all quadrature points, which in the SEM framework coincide with the nodal points whose values need to be computed.
The final step of the computation involves calculating each point’s residual contribution for the current term \circlednum{3}. Since this step updates the global residual array, it encompasses multiple atomic updates to ensure proper memory synchronization, which serializes memory accesses and updates. 

We consider analysis on two cases for turbulence simulations: a TGV (Taylor-Green Vortex) and a CF (Channel Flow) case configuration.  Performance profiling reveals consistent hotspot behavior across both simulated cases, with the \code{full\_convec} (Convection) kernel dominating execution time -accounting for up to 50\% (NVIDIA) and 70\% (AMD) of total runtime— followed by the \code{full\_diffusion} (Diffusion) kernel at 22\% and 10\% respectively. Given the strong similarity in hotspot distribution, subsequent analysis focuses on the Channel Flow (CF) case as more representative to air-flow in urban environments, which describes the flow of fluid confined between two parallel surfaces/walls, typically forming a channel. The Convection kernel remains the primary computational bottleneck, showing notable sensitivity to floating-point precision and GPU vendor and mesh size, denoting the number of points in the unstructured mesh.


\section{Multi-Level Design Space Exploration}
\begin{figure}[t]
  \centering
  \includegraphics[width=1\columnwidth]{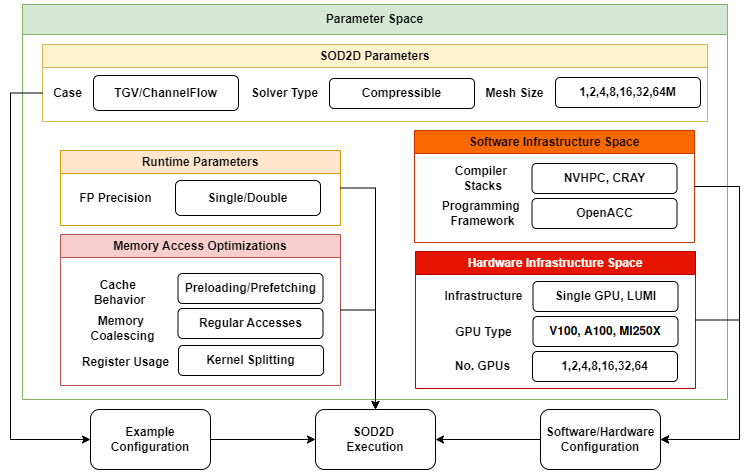}
  \caption{Organization of the examined multi-level design space.}
  \label{fig:parameter_space}
  \vspace{-14pt}
\end{figure}

As illustrated in Fig. \ref{fig:parameter_space} we examine the application's behavior under all parameter space layers. We start from the configuration of the application's input parameters, extending down to the hardware infrastructure parameters, while also exploring code memory optimizations.
The \textit{Runtime Parameters layer} of our exploration framework considers single (FP32) and double (FP64) precision simulations. The Memory Access Optimizations section hierarchically fits between the Runtime Parameters and the Software Infrastructure space as it refers to core, manual code modifications elaborated on in subsection \ref{sec:mao}. 
 The \textit{Software Infrastructure layer} includes the compiler frameworks under examination, specifically NVHPC v24.5 for our local NVIDIA GPU and Cray Clang v17.0.1 for the LUMI cluster, utilizing AMD GPUs. 
With respect to the \textit{Hardware Infrastructure layer}, a distinction between single and multi-GPU runs is made, as well as between different accelerator vendors.

Interestingly, initial profiling results demonstrated a 15.83$\times$ slowdown of a single AMD MI250X Graphics Computing Die (GCD) compared to the NVIDIA Tesla V100 for ~8M nodes, with single-precision arithmetic and without any code optimizations enabled.
We attribute these disparities to four reasons: i) to the power cap on the GPUs of LUMI being just 400W compared to their nominal value of 500W, ii) to the use of a single GCD of the MI250X, iii) to the potentially unoptimized Cray compiler stack w.r.t. to  the underlying hardware, iv) to the single-precision throughput of the MI250X faring much worse against that of the V100 compared to double-precision.

\subsection{Memory Access Optimizations}
\label{sec:mao}
 
Profiling results of the application on multi-GPU runs with Rocprofv3 and Omnitrace revealed maxed out local memory usage, along with register usage and inefficient coalescing of memory accesses in both shared and global memory. To this end, we parameterized the target kernels with memory access optimization strategies, applied with hand-crafted templates to the source code to alleviate each of the aforementioned issues. Details on the memory access optimizations follow:

\textbf{Kernel Splitting:} \code{full\_convec}, the main hotspot of SOD2D, executes the steps \circlednum{1}-\circlednum{3} of Fig. \ref{fig:SOD2D-flow} for Mass, Momentum and Energy equations concurrently, fetching global subarrays to per-element private buffers.
The aforementioned steps are independent with regards to the partial results computed and the final output memory location.
Due to the increased per-thread register pressure on the MI250X, {\it i.e. Average VGPR being equal to Maximum at 128, Average LDS Allocation being equal to Maximum at 4608} we opted to split the kernel into the three separate kernels corresponding to each equation. This provided an immediate 1.5$\times$speedup on \code{full\_convec}.

\textbf{Preloading:}
\label{subsec:preloading}
The preloading strategy is commonly employed in GPU-optimized codes to (1) improve data spatial locality, (2) potentially take advantage of shared memory within thread blocks, and (3) maximize memory bandwidth by coalescing memory requests to fetch multiple elements simultaneously.

The default version of SOD2D employs this technique \cite{GASPARINO2024109067}, as illustrated in stage \circlednum{1} of Fig. \ref{fig:SOD2D-flow}. It is applied to global matrices that need to be accessed with an element index and are initially stored in an irregular manner. Since these matrices are in the Global Memory Space of the GPU and are accessed frequently in the Execution stage (\circlednum{2a}-\circlednum{2c}) of Fig. \ref{fig:SOD2D-flow}, they are first fetched to element-private local variables which removes the element node index, making memory accesses more regular and closer to computation. 

Removing these preloaded variables and accessing global matrices directly provides moderate performance gains on many types of runs. We have attributed this effect to the sheer size of the local matrices, which leads to ineffective utilization of the lowest parts of the memory hierarchy.

 \textbf{Prefetching:}
Prefetching refers to the use of the openacc \code{cache} directive at specific points in the \code{full\_convec} subroutine. According to the OpenACC Standard 2.7, those directives are tied to loops located above or inside of and specify elements or ranges of subarrays that should be fetched into the highest level of the cache for the loop body.

Intuitively, the prefetching mechanism cannot be effective when applied to large global arrays with indirect access patterns. Range-based prefetching of global arrays was found to be ineffective. 
Therefore, our utilization of this directive is coupled with the preloaded local arrays described in subsection \ref{subsec:preloading}.  Moreover, prefetching the local arrays for every quantity in the quadrature points loops is also ineffective. We have concluded that the optimal usage of this directive regards the quantities that participate in the deepest nested loops in the body of the quadrature loop. This provides a 1.24$\times$ speedup with reference to just Preloading for the CF case with FP64, the split kernel and the V100 GPU.

\begin{figure*}[t!]
  \centering
  \includegraphics[width=\linewidth]{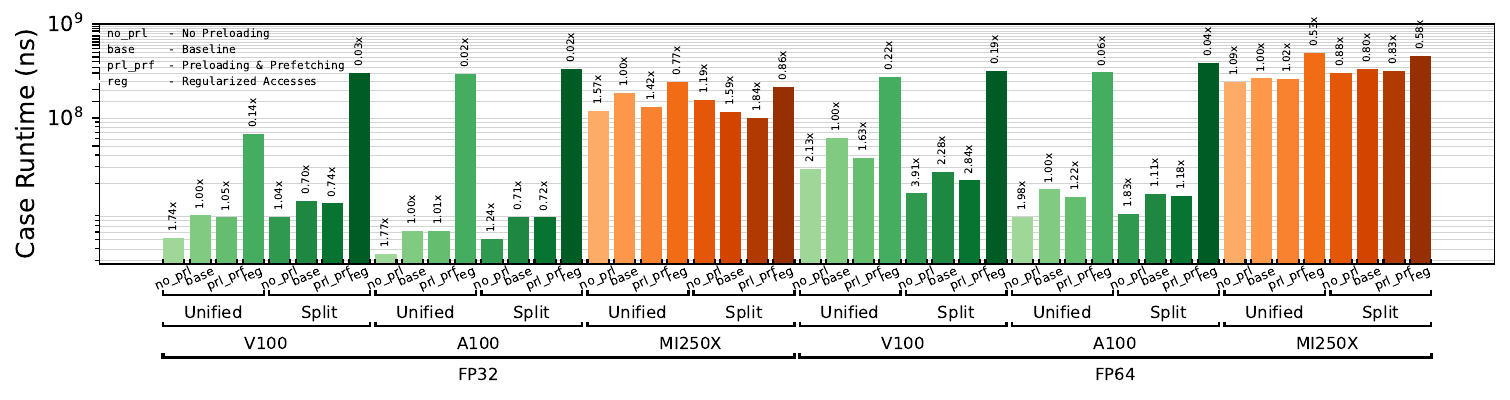}
  \caption{Single-GPU run times across all kernel optimizations for Channel Flow.}
  \label{fig:combined_figures}
\end{figure*}

\textbf{Regular Accesses:}
A common way to improve GPU memory bandwidth is memory coalescing. Since sub-optimal coalescing ({\it 6.5 sectors/request, max 16}) limits performance, we regularized global memory accesses in the ``No Preloading'' variant of SOD2D. In the quadrature loop (\circlednum{2a}-\circlednum{2c} in Fig.~\ref{fig:SOD2D-flow}), accesses follow \code{rho(connec(ielem,invAtoIJK()))}, where \code{connec} is the mesh connectivity matrix and \code{invAtoIJK} an index linearization function. Such macros cause major performance penalties~\cite{SATHYANARAYANA2025104993}. As \code{connec} and \code{invAtoIJK} depend only on static mesh topology, we precompute \code{connec(ielem,invAtoIJK())} and store it in a global matrix indexed by element node, quadrature point, and polynomial order. Subsequent kernel launches reuse it with perfect memory coalescing. Although \code{rho} accesses remain irregular, the costly linearization is removed. The matrices must fit in lower memory levels for efficiency.

\section{Evaluation}
\label{sec:eval}


\subsection{ Experimental Setup}
In this section, we present the results of our evaluation, providing insights on the behavior of SOD2D under different configurations of the parameter space. We examine the CF case with a mesh size of 1, 2, 4, 8, 16 million nodes with compressible solver. All input cases were run on both FP32 and FP64 modes of SOD2D.
Regarding the SOD2D versioning, the application was compiled a total of 8 times. More specifically, the versions were SOD2D without preloading (\code{\bf no\_prl}), with preloading (\code{\bf base}), preloading and prefetching (\code{\bf prl\_prf}) and regular memory accesses (\code{\bf reg}), each of which were with (\code{\bf split}) and without (\code{\bf unified}) kernel splitting. 

All cases and SOD2D versions were run on both NVIDIA and AMD systems. Specifically, a single NVIDIA Tesla V100-PCIE-32GB GPU in a local server as well as an NVIDIA A100 GPU from the Berzelius SuperPOD and a single Graphics Compute Die (GCD) of the AMD CDNA™ 2 MI250X-128GB courtesy of LUMI HPC cluster were utilized. Scalability experiments were run exclusively on LUMI.

\subsection{Single-GPU Exploration}
Since near identical results are obtained between the TGV and CF cases in this section, we will focus on the more realistic CF case.
The multi-level parameter space exploration and Cross-vendor single GPU comparative analysis reveals varying behavior of the main hotspot (\code{full\_convec}) when sweeping different parameters, highlighting the potential for the selection of locally optimal modifications. 

As mentioned previously, the slowdown seen on a single GCD of the MI250X compared to the V100 is 15.83$\times$ on 8M nodes and FP32 (23.38$\times$ for the A100). The respective slowdown value for the split kernel is 9.87$\times$ and 8.66$\times$ for the \code{\bf prl\_prf} version of SOD2D. For double-precision, the latter value is 5.23$\times$. Regarding performance disparity between NVIDIA GPUs, the A100 provides a 1.47$\times$ and a 3.05$\times$ speedup over the V100 on FP32 and FP64 respectively.

Since AMD is consistently outperformed, the relative slowdowns compared  to NVIDIA are not noted on Fig. \ref{fig:combined_figures}. Instead, runtimes are evaluated across the same vendor and FP precision. Each bar corresponds to the average of 5 runs ranging from 1 to 16 million nodes, therefore the plot corresponds to a total of 240 runs of SOD2D.

We note that baseline SOD2D comes optimized with the Preloading technique detailed in Section~\ref{sec:mao}. This process is expensive in terms of runtime since the program flow must go through the same loop bounds as \circlednum{2a}-\circlednum{2c} to preload global variables. Since inherent prefetching to caches in the \code{\bf no\_prl} version can be equally efficient, it is not obvious whether the benefit of Preloading compensates for the overhead of block \circlednum{1}. As depicted in Fig. \ref{fig:combined_figures}a, the \code{\bf no\_prl} version always outperforms the \code{\bf base} (Preloading) version, with the exception of the split version with FP64 on the MI250X. This effect is greater for FP64 (DP) than FP32 (SP) and more prominent for the AMD MI250X compared to NVIDIA GPUs for FP32, while the opposite is true for FP64.

Preloading paired with Prefetching (\code{\bf prl\_prf}) consistently improves its performance (\code{\bf base}). The effect is mostly minor, except for AMD MI250X on FP32, and NVIDIA V100 on FP64 regarding the split kernel. In the former case, a 42\% and 16\% speedup is provided for unified and split kernel respectively, while in the latter, a 63\% and 25\% speedup is observed, rendering it the optimal code modification for AMD on FP32 and NVIDIA on FP64. We once again note the alternating effect of optimizations between NVIDIA-AMD and FP32-FP64.

Regular Accesses (\code{\bf reg}) underperform on the different platforms due to the size of the regular matrix being 18$\times$ that of \code{connec}, exceeding the capacity of the lowest parts of the memory hierarchy. Thus, irregular accesses to a smaller matrix are favorable to regular accesses to a larger one.  However, \code{\bf reg} is the only parameter that displays drastically different behavior across the cases run. Specifically for FP32, it fares better on the unified kernel. Once again, this behavior alternates on AMD. Such minor differences are consistent for FP64.

The impact of Kernel Splitting (\code{\bf split}) is highly variable across vendors and floating point precisions as well, but mostly consistent across cases. For FP32, it speeds up the execution on AMD but incurs a slowdown for NVIDIA. Interestingly, the opposite behavior is displayed for FP64.

The behavior outlined above is consistent across all optimizations except Access Regularization.  We attribute these differences to the differently sized Register Files for FP32 and FP64 across architectures.


\subsection{Scalability Analysis}
\begin{figure}[t]
    \centering
    \includegraphics[width=1.\columnwidth]{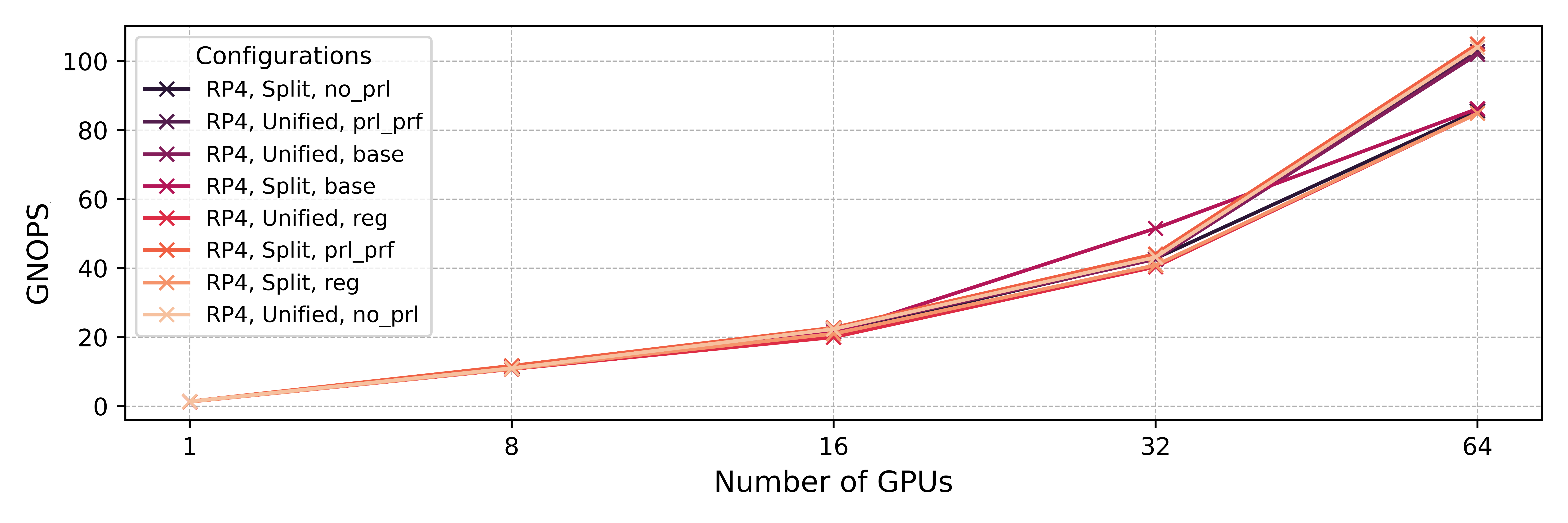}
    \caption{Throughput at scale for Channel Flow.}
    \label{fig:channelflowatscale}
\end{figure}

Fig.~\ref{fig:channelflowatscale} represents data that resulted from 160 runs of a weak scaling problem. For each configuration of the total number of GPUs, the size of the problem in terms of  Nodes increases linearly, providing the same amount work per GPU, which corresponds to approximately 1M Nodes.

In the context of scalability analysis of SOD2D, we evaluate efficiency as a metric of throughput. Thus, we introduce the \textit{Giga Node Operations Per Second (GNOPS)} as:

\textit{$$
\frac{\text{\#TotalMeshNodes}\times \text{\#RungeKuttaSteps}\times  \text{\#TotalTimeSteps}} {\text{SOD2DRuntime(ns)}}
$$}
\noindent \textit{GNOPS} metric emerges from the homogeneous way the nodes are updated throughout the steps \circlednum{1}-\circlednum{3} in Fig.~\ref{fig:SOD2D-flow}.

As observed in Fig.~\ref{fig:channelflowatscale} that shows runs on FP32 (RP4), scaling is mostly consistent for all of the optimizations (FP32 displays 1.22$\times$ higher GNOPS than FP64 on average overall.). However, we observe that CF demonstrates high inconsistency between scaling of different memory optimizations, with optimal configurations alternating between different scales (superlinear performance is observed due to partitioning inefficiencies at specific numbers of nodes \cite{date24}). We report that for the Unified kernel with no Preloading (\code{\bf no\_prl}) is the optimal for all GPU configurations. We attribute this to redundant preloading operations. For CF the improvements from worst to optimal configurations are 23.8\% (FP32) and 14.6\% (FP64) respectively. 
Therefore, as is apparent in Fig. \ref{fig:channelflowatscale}, selecting an optimization for Channel Flow at 64 GPUs based on the optimum at 32 GPUs, would incur a 23.8\% (FP32) and 8.3\% (FP64) throughput penalty. Such disparities are absent in TGV, with 12.3\% variation and consistent optimal configurations. The above analysis provides an important lesson for large scale accelerated compressible fluid dynamics, that \textit{on a case-independent basis, performance prediction at different scales through simply extrapolating on low scale optimal memory access decisions is precarious, i.e. more sophisticated modelling at scale is required}.



\section{Conclusions}
    
This paper explored the performance portability challenges in GPU-accelerated CFD simulations through a multi-level analysis of the SOD2D framework. 
We highlighted how architectural differences between GPU vendors significantly impact solver performance,   revealing substantial disparities in scalability and performance portability, underscoring the need for parameter-space-informed optimizations. This study highlights the importance of multi-level exploration for developing performance-portable CFD frameworks, enabling efficient adaptation to heterogeneous HPC environments. Future work will extend this approach to new compiler and hardware infrastructure as well as other types of simulations and input meshes with an emphasis on further automating parameter optimization.








{
\bibliographystyle{IEEEtran}
\bibliography{sample-base}
}

\vspace{12pt}

\end{document}